\documentclass[prb,twocolumn,showpacs,preprintnumbers,amsmath,amssymb,superscriptaddress,nofootinbib]{revtex4}
\pdfoutput=1

\usepackage{bbm,bm,slashed}
\usepackage{graphicx,graphics,psfrag,epsfig}
\usepackage{epstopdf}
\usepackage{color,array,tabularx}
\usepackage{hyperref}
\usepackage{soul}
\usepackage[T1]{fontenc}

\newcommand{\be}{\begin{eqnarray}}
\newcommand{\ee}{\end{eqnarray}}

\newcommand{\nn}{\nonumber } 
\newcommand{\Eqref}[1]{Eq.~\eqref{#1}}

\begin{document}

\author{Daniel D. Scherer}\email{daniel.scherer@nbi.ku.dk}
\affiliation{Niels Bohr Institute, University of Copenhagen, DK-2100 Copenhagen, Denmark}
\affiliation{Institut f\"ur Theoretische Physik, Universit\"at Leipzig, D-04103 Leipzig, Germany}

\author{Michael M. Scherer}
\affiliation{Institut f\"ur Theoretische Physik, Universit\"at
Heidelberg, D-69120 Heidelberg, Germany}

\author{Carsten Honerkamp}
\affiliation{Institute for Theoretical Solid State Physics, RWTH Aachen University, D-52056 Aachen, Germany
and JARA - FIT Fundamentals of Future Information Technology, Germany}

\title{Correlated spinless fermions on the honeycomb lattice revisited}

\begin{abstract}
We investigate the quantum many-body instabilities of the extended Hubbard model for spinless fermions on the honeycomb lattice with repulsive nearest-neighbor and 2nd nearest-neighbor density-density interactions. Recent exact diagonalization and infinite density matrix renormalization group results suggest that a putative topological Mott insulator phase driven by the 2nd nearest-neighbor repulsion is suppressed, while other numerically exact approaches support the topological Mott insulator scenario. In the present work, we employ the functional renormalization group (fRG) for correlated fermionic systems. Our fRG results hint at a strong suppression of the scattering processes stabilizing the topological Mott insulator. From analyzing the effects of fermionic fluctuations, we obtain a phase diagram which is the result of the competition of various charge ordering instabilities.
\end{abstract}

\maketitle

\section{Introduction}
\label{sec:intro}

The extended Hubbard model for spinless fermions at half-filling is possibly the simplest itinerant fermion model with interactions to be put on the honeycomb lattice. Yet, it features interesting interaction driven effects as, e.g., the possible realization of an interaction induced \textit{topological Mott insulator}~\cite{raghu2008} with a symmetry-protected chiral edge state.

The effect of a repulsive nearest-neighbor interaction $V_{1}$ is comparatively well-understood: Beyond a critical coupling strength $V_{1}$ destabilizes the semi-metallic regime and the system undergoes a direct and continuous strong coupling quantum phase transition (QPT) to a fully gapped charge-density wave state (CDW). The critical exponents~\cite{herbut2006,herbut2009,rosa2001,hofling2002,braun2011} obtained for this quantum critical point (QCP) show that it falls into the Gross-Neveu~\cite{gross1974,rosenstein1988} universality class\footnote[2]{We note that the critical exponents of Gross-Neveu type quantum field theories depend in fact on flavor number and also the number of spinor components. Here, we are dealing with two flavors of 4-component Dirac fermions, corresponding to a reducible representation of the Dirac algebra in $2+1$ spacetime dimensions. This universality class is also referred to as Gross-Neveu-Yukawa with $\mathbbm{Z}_{2}$ order parameter or chiral Ising universality class~\cite{rosenstein1993,janssen2014,classen2015}.}. The QPT corresponding to CDW order was investigated by a variety of methods, ranging from a Dyson-Schwinger equation based analysis~\cite{khveshchenko2001} to non-perturbative renormalization group calculations~\cite{rosa2001,hofling2002,braun2011} and recent Quantum Monte Carlo~\cite{wang2014} and newly developed Majorana Quantum Monte Carlo simulations~\cite{li2015A,li2015B}.

Considering only a repulsive 2nd nearest-neighbor interaction $V_{2}$, the situation presently appears to be far from clear. Previous mean-field theories~\cite{raghu2008,weeks2010,dauphin2012,grushin2013} found a direct and continuous, strong coupling QPT into a symmetry broken phase with a complex bond-order (BO) parameter $\chi_{ij}$, where $ij$ denotes a 2nd nearest-neighbor pair. On a mean-field level, a finite dimerization $\chi_{ij}$ on a 2nd nearest-neighbor bond renormalizes the bare hopping matrix and breaks inversion symmetry $\mathcal{P}$ (real part of $\chi_{ij}$) and/or time-reversal symmetry $\mathcal{T}$ (imaginary part of $\chi_{ij}$). It was found~\cite{raghu2008} that the mean-field ground-state energy is minimized by the solution breaking time-reversal symmetry only. This peculiar state would be an interaction induced realization of the Haldane model~\cite{haldane1988}. The latter is a topologically non-trivial model for non-interacting spinless fermions on the honeycomb with nearest and 2nd nearest-neighbor hopping. The breaking of time-reversal symmetry is due to a particular flux configuration, which does not, however, correspond to a finite homogeneous magnetic field penetrating the system. The two band model features Chern numbers $\pm 1$ and, correspondingly, for a full bulk gap a quantized Hall conductivity, which is carried by a gapless chiral edge state at the sample boundary. This state is therefore referred to as a \textit{quantum anomalous Hall} state: it is characterized by a topologically protected chiral edge state, but as opposed to the integer quantum Hall state, does not require a net magnetic flux through the sample. In the sense of the Altland-Zirnbauer classification of topological insulators and topological superconductors, the Haldane model falls into the unitary symmetry class A~\cite{ryu2010}. Therefore, in Ref.~\onlinecite{raghu2008}, the mean-field state for the \textit{interacting} spinless fermion model on the honeycomb lattice was dubbed topological Mott insulator.

Numerical studies of spinless fermions on the honeycomb and the spinless $\pi$-flux model on the square lattice report the absence of an interaction induced quantum anomalous Hall (QAH) phase. In Refs.~\onlinecite{jia2013,daghofer2014,capponi2015,motruk2015} the observation of a direct transition from the semi-metallic (SM) phase to a modulated charge density wave phase driven by repulsive 2nd nearest neighbor density-density interaction was reported.
In this work, we follow Ref.~\onlinecite{mmscherer2012,mmscherer2012b} and call this particular type of modulated charge density wave phase a \textit{three-sublattice charge density wave}, CDW$_{3}$ for short\footnote{Note that in Refs.~\onlinecite{jia2013,grushin2013,garcia2013,daghofer2014,capponi2015,motruk2015} the abbreviation for this charge-modulated insulating state is CM(s).}. In Ref.~\onlinecite{daghofer2014}, the authors infer from cluster perturbation theory that while QAH correlations exist for small clusters, the QAH state ceases to be the ground-state for increasing cluster size. The lack of a long-range ordered QAH state is further attributed to the vanishing density of states at half-filling, which leads to an insufficient energy-gain from the formation of a QAH dimerization pattern. Another exact diagonalization study~\cite{garcia2013} finds an intermittent Kekul$\acute{\textrm{e}}$ dimerization phase, sandwiched between SM and modulated charge density phase, but with the QAH state also completely squeezed out of the phase diagram. The Kekul$\acute{\textrm{e}}$ dimerization phase was recently corroborated beyond mean-field theory in the strong-coupling regime by ED studies on large clusters~\cite{capponi2015} and infinite density matrix renormalization group (iDMRG)~\cite{motruk2015}. These works further revealed charge-ordered ground-states not seen in previous mean-field phase diagrams.

These results represent drastic revisions of the mean-field phase diagrams~\cite{raghu2008,weeks2010,dauphin2012,grushin2013} for spinless fermions at half-filling with Dirac-type low-energy excitations. These findings might further indicate the presence of strong fermionic and possibly collective bosonic quantum fluctuations close to the would-be transition from the semi-metallic state to the QAH state. The precise nature of these strong quantum fluctuations, however, remains somewhat elusive. The SM to QAH mean-field transition does not break a continuous symmetry~\cite{raghu2008}, ruling out the reduction of possible ordering tendencies in a given channel due to the backaction of gapless collective degrees of freedom. But the ground-state reported by exact diagonalization for $V_{1} = 0$, $V_{2}>0$ comes with a huge degeneracy in the classical limit. On the quantum level, there is, however, only a small finite subset of degenerate ground states with different charge arrangements. Since only discrete symmetries are broken, also in the case of the CDW$_{3}$, we do not expect a strong influence of collective fluctuations.

We note, however, that other numerically exact and variational approaches employed in Ref.~\onlinecite{duric2014} do find support for the interaction induced QAH state. The differences between Refs.~\onlinecite{jia2013,daghofer2014} and~\onlinecite{duric2014} concern system sizes and boundary conditions: periodic~\cite{jia2013,daghofer2014} vs. open~\cite{duric2014}. In our functional renormalization group approach to the problem, we work with an infinite system with periodic boundary conditions.

In the following, we will present our results obtained within the functional renormalization group (fRG) framework for correlated fermions.
One of our main results in the present work is that the unbiased inclusion of fermionic fluctuations with a refined resolution of momentum space is sufficient, to obtain the CDW$_{3}$ with a finite wavevector transfer $\vec{Q}$ as the leading instability from our renormalization group flows, as the interaction strength $V_{2}$ is increased to destabilize the semi-metal. We thus provide further evidence that the sought after topological Mott insulator phase for spinless fermions on the honeycomb is destroyed by competing fermionic fluctuations in the particle-hole channel, and is replaced by a fully gapped, charge-ordered state with an enlarged unit cell. We further investigate the phase diagram for both $V_{1},\, V_{2} > 0$.

The present paper is structured as follows. In Sect.~\ref{sec:model}, we introduce the Hamiltonian for spinless fermions on the honeycomb lattice with repulsive nearest and 2nd nearest-neighbor interactions. In Sect.~\ref{sec:frg}, we recap the essentials of the fRG method we employ to analyze the phase diagram of the model. In Sect.~\ref{sec:instabilities}, we present our results for the phase diagram, as the strengths of both nearest and 2nd nearest-neighbor interactions are varied. Sect.~\ref{sec:suppression} is devoted to a detailed investigation of the relevant scattering processes which lead to a suppression of the QAH instability and ultimately favor ordering tendencies corresponding to a charge modulated ground state (CDW$_3$).

\section{The Model}
\label{sec:model}

%
\begin{figure}[t!]
\begin{minipage}{0.35\columnwidth}
\centering
\includegraphics[width=1.0\columnwidth]{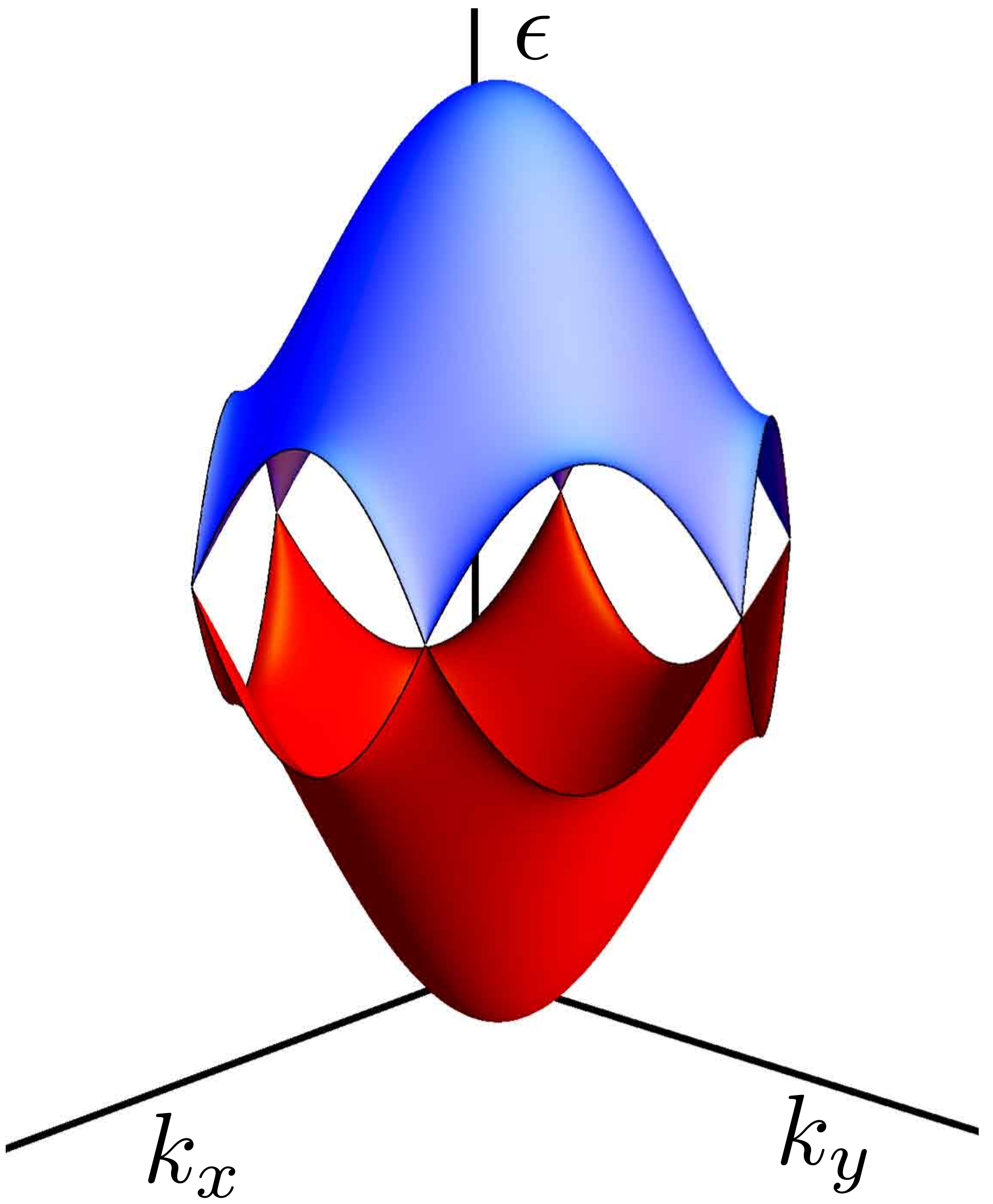}
\end{minipage}
\hspace{0.5em}
\begin{minipage}{0.5\columnwidth}
\centering
\includegraphics[width=1.0\columnwidth]{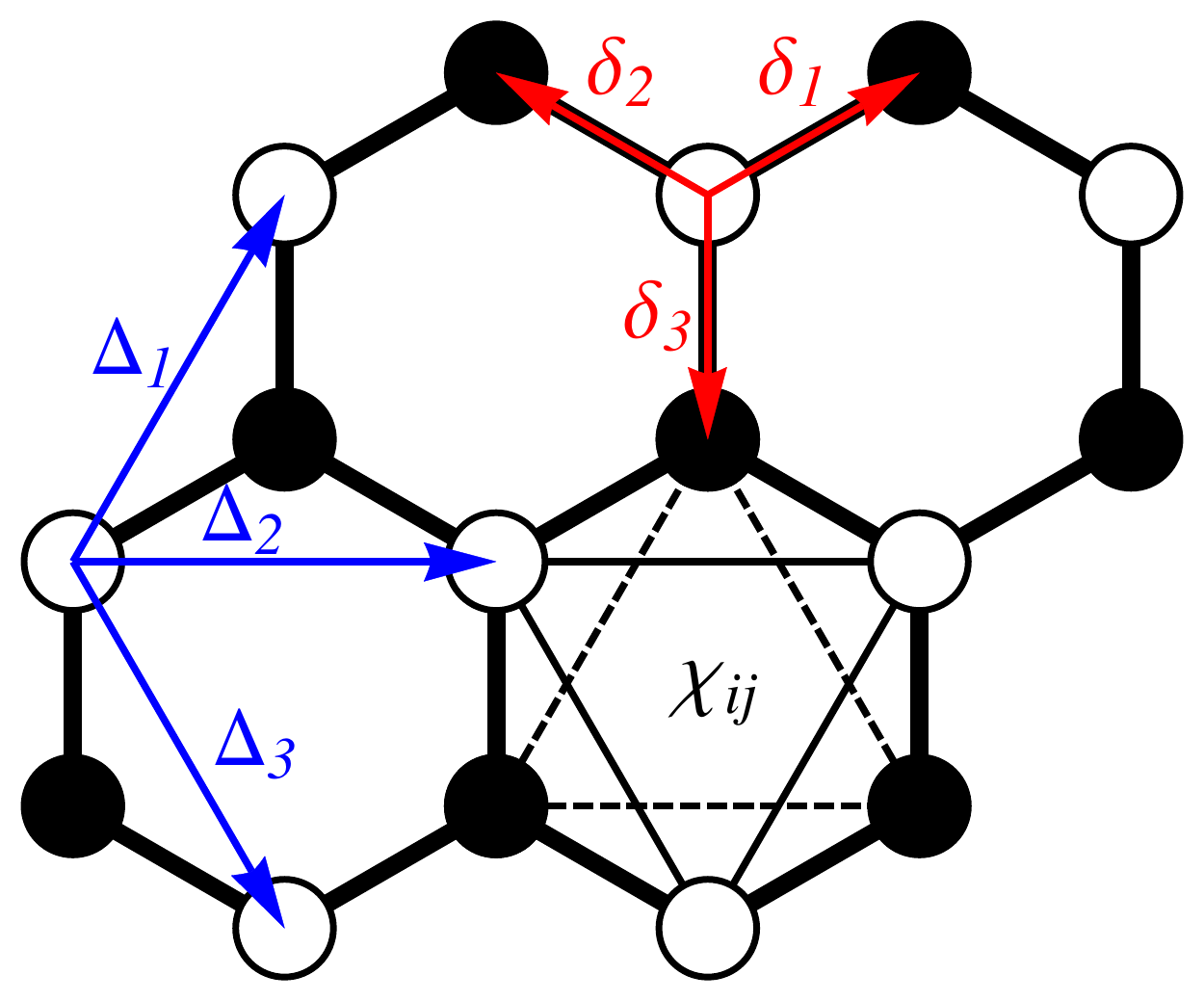}
\end{minipage}
\caption{Left panel: Energy bands for spinless fermions on the bipartite honeycomb lattice with nearest-neighbor hopping amplitude $t$.
For the system at half-filling, the valence band (red) is completely filled while the conduction band (blue) is completely empty. The dispersion is approximately linear around the $K$ and $K^{\prime}$ points. Right panel: Lattice geometry and bond dimerization $\chi_{ij}$, cf. Ref.~\onlinecite{raghu2008}. White disks correspond to sites of the $A$ sublattice, black disks correspond to those of the $B$ sublattice. Red arrows correspond to nearest-neighbor vectors $\vec{\delta}_{1}$, $\vec{\delta}_{2}$, $\vec{\delta}_{3}$, and blue arrows correspond to 2nd nearest-neighbor vectors $\vec{\Delta}_{1}$, $\vec{\Delta}_{2}$, $\vec{\Delta}_{3}$. Thin black lines (solid: $A$ sublattice, dashed: $B$ sublattice) visualize the bond order $\chi_{ij}$.}
\label{fig:model}
\end{figure}
The Hamiltonian $H$ of the spinless fermion model on the honeycomb is decomposed into a hopping part, $H_{0}$, and an interaction part $H_{\mathrm{int}}$,
\be
H =  H_{0} + H_{\mathrm{int}}.
\label{eq:H}
\ee
In the following, we only consider nearest-neighbor hopping with a real hopping amplitude $t$,
\be
H_{0}  & = & -t\sum_{\langle i ,j \rangle} \left(c_{i}^{\dagger} c_{j} + c_{j}^{\dagger} c_{i} \right) + \mu \sum_{i}c_{i}^{\dagger}c_{i}.
\label{eq:H0}
\ee
The interaction term reads
\be
H_{\mathrm{int}} & = & V_{1}\sum_{\langle i , j\rangle} n_{i} n_{j} + V_{2}\sum_{\langle\!\langle i , j \rangle \! \rangle} n_{i} n_{j}.
\label{eq:Hint}
\ee
Here, the operator $c_{i}^{\dagger}$ creates a spinless fermion at lattice site $i$. The symbol $\sum_{\langle i,j \rangle}$ denotes a sum over nearest-neighbor pairs of lattice sites connected by the vectors $\vec{\delta}_{i}$, $i=1,2,3$, where each pair is counted only once. Analogously, $\sum_{\langle\!\langle i , j \rangle\!\rangle}$ denotes a sum over 2nd nearest-neighbor pairs connected by the vectors $\vec{\Delta}_{i}$, $i=1,\dots,6$, where again each pair is counted only once. See Fig.~\ref{fig:model} for a depiction of the lattice geometry.

Due to the two sublattices with inequivalent nearest-neighbor sites,
the hopping Hamiltonian gives rise to a two-band model. The half-filling condition, which we will enforce througout the rest of the paper, simply translates to a vanishing chemical potential, $\mu = 0$. For vanishing interactions, the simple nearest-neighbor hopping leads to a band structure analogous to that of the tight-binding model for graphene. The Fermi energy is poised at the Dirac points $K$, $K^{\prime}$. The fermionic low-energy excitations feature a linear dispersion around these high-symmetry points in the Brillouin zone, cf. Fig.~\ref{fig:model}, where the valence and conduction bands touch. Correspondingly, the non-interacting single-particle density of states vanishes, as the energy of the non-interacting fermionic degrees of freedom approaches the Fermi energy. Without interactions, the model can be understood as a semi-metal.
The vanishing of the non-interacting single-particle density of states renders the semi-metallic state stable with respect to correlation effects, such as the opening of a gap in the single-particle spectral function, or possible symmetry breaking quantum phase transitions.

We note that for spinless fermions, the half-filling condition implies one fermion per 2-atom unit cell.
The interaction induced quantum anomalous Hall state characterized by a complex bond-order field $\chi_{ij}=\langle c_{i}^{\dagger} c_{j}\rangle$ is described in good detail in Ref.~\onlinecite{raghu2008}.

\section{Functional renormalization group essentials}
\label{sec:frg}

In this work, we employ a functional renormalization group approach for the one-particle-irreducible (1PI) vertices~\cite{negorl} with an energy cutoff. For  recent reviews on the fRG method, see Refs.~\onlinecite{kopietz2010,metzner2011,platt2013}.
The fRG calculation is most economically performed in the band-basis, which diagonalizes the quadratic part $H_{0}$ of the fermion Hamiltonian. The fRG equations can be derived from the functional integral representation of the fermionic partition function.
\begin{figure}[t!]
\centering
\includegraphics[width=0.85\columnwidth]{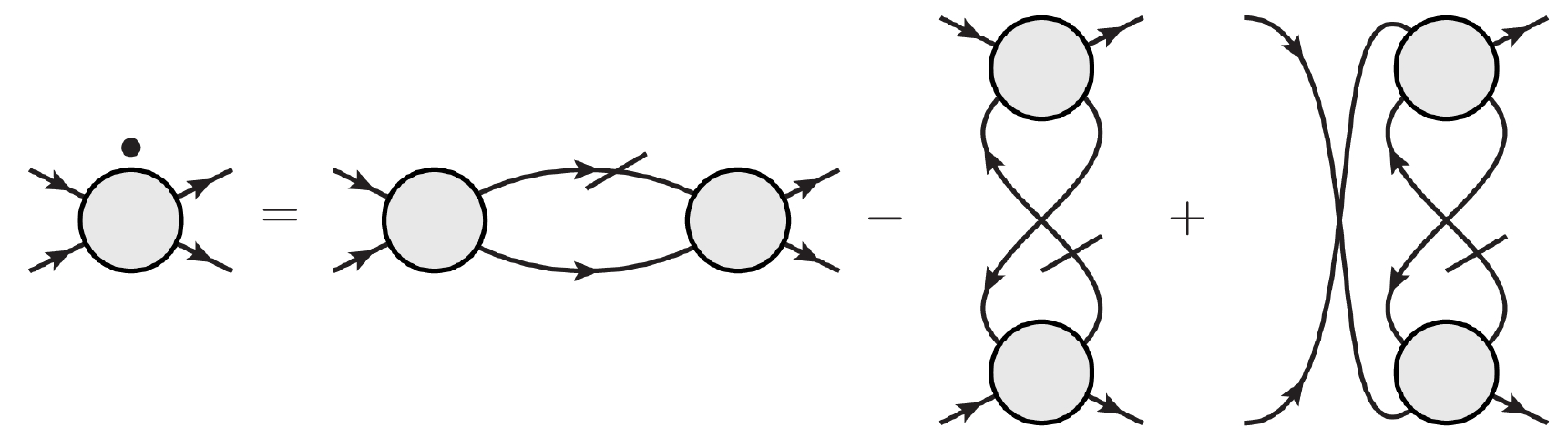}
\caption{Diagrammatic representation of the functional renormalization group equation
for the scale-dependent 4-point vertex $V^{\Lambda}$. The grey filled circle denotes $V^{\Lambda}$, while the
black dot on the left hand side corresponds to the scale-derivative $\frac{d}{d\Lambda} V^{\Lambda}$. Internal lines correspond
to the fermion propagator $G^{\Lambda}$, while slashed lines correspond to the single-scale propagator $S^{\Lambda}$. Diagrams
with slashed and unslashed lines exchanged are not shown for brevity. The first diagram on the right hand side corresponds to 
fermion fluctuations in the particle-particle channel. The remaining two diagrams encode the fluctuations in the particle-hole channel.}
\label{fig:diagrams}
\end{figure}
The corresponding Matsubara action of the model~\Eqref{eq:H} can be compactly written as
\be
S[\bar{\psi},\psi] = -(\bar{\psi},G_{0}^{-1}\psi) + V[\bar{\psi},\psi].
\ee
In the band representation, where the fermion fields correspond to the eigenstates of the hopping Hamiltonian, the bilinear part becomes
\be
(\bar{\psi},G_{0}^{-1}\psi)=\sum_{b,k}\bar{\psi}_{b}(k)\left(\mathrm{i}\omega_{n} - \epsilon_{b}(\vec{k})\right)\psi_{b}(k),
\ee
where $k = (\mathrm{i}\omega_{n},\vec{k})$ and the fermion fields are labeled by the band index $b=v,c$ ($v$: valence, $c$: conduction band).

The interaction functional $V[\bar{\psi},\psi]$ reads
\be
V[\bar{\psi},\psi] & = & \frac{1}{4}\sum_{\{b_{i}\},\{k_{i}\}}V_{b_1,b_2,b_3,b_4}(k_1,k_2,k_3,k_4)\times \nn \\
& & \bar{\psi}_{b_1}(k_1)\bar{\psi}_{b_2}(k_2)\psi_{b_3}(k_3)\psi_{b_4}(k_4).
\ee
The properly antisymmetrized coupling function $V_{b_1,b_2,b_3,b_4}(k_1,k_2,k_3,k_4)$ is obtained from the interaction Hamiltonian \Eqref{eq:Hint} by substituting operators by Grassmann fields and going from the tight-binding representation to the band-representation. In the process, the coupling function picks up an additional momentum dependence -- so-called orbital make-up (see e.g. Ref.~\onlinecite{maier2013}) -- due to the non-trivial sublattice structure of the hopping Hamiltonian. The frequency dependence is simple and just encodes the Matsubara frequency conservation of the instantaneous interaction.
\begin{figure}[t!]
\centering
\includegraphics[height=0.6\columnwidth]{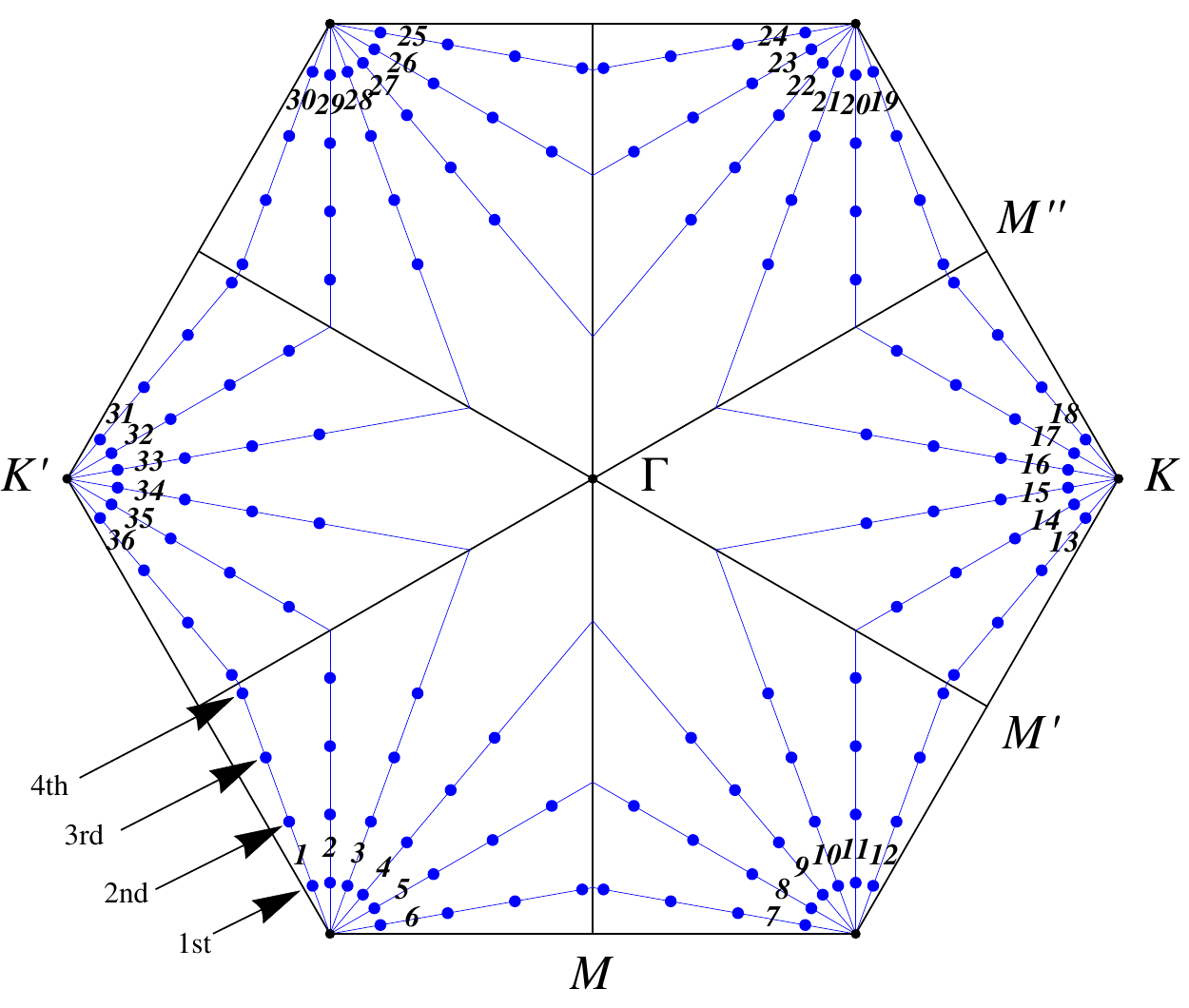}
\caption{The figure illustrates the discretization scheme for the momentum  dependence for the external legs of the vertex function in the first Brillouin zone for a total of $N=144$ patch points (blue points). The momenta $\vec{k}_{1}$, $\vec{k}_{2}$, $\vec{k}_{3}$ are projected onto the patch points, denoted by $\pi(\vec{k})$. These are chosen to lie on patch rings around the $K$, $K^{\prime}$ points. We refer to the patch ring closest to $K$ or $K^{\prime}$ as the first patch ring, the next largest radius corresponds to the second patch ring, and so on. See Appendix~\ref{app:patching} for details of the projection scheme. Due to translation invariance, the fourth momentum $\vec{k}_{4}$ is projected onto the patch point dictated by momentum conservation. We evaluate the flow equations~\Eqref{eq:flow} for the projected vertex function $V^{\Lambda}(\pi(\vec{k}_{1}),\pi(\vec{k}_{2}),\pi(\vec{k}_{3}),\pi(\vec{k}_{4}))$. The scale-resolved loop integrations are discretized along the blue lines emanating from the $K$, $K^{\prime}$ points in radial direction. For clarity, we only enumerate the 36 patch points closest to the $K$, $K^{\prime}$ points. The patch rings further away from the $K$, $K^{\prime}$ points are labeled analogously.}
\label{fig:patching}
\end{figure}
In the 1PI fRG-scheme, an infrared regulator with energy scale
$\Lambda$ regularizes the bilinear contribution of the bare action in the
functional integral. The \textit{regularized} bare propagator replaces the bare propagator according to
\be
G_0(b,\mathrm{i}\omega_{n},\vec{k})\rightarrow
G_0^\Lambda(b,\mathrm{i}\omega_{n},\vec{k})=\frac{C^\Lambda
[\epsilon_{b}(\vec{k})] }{i \omega_{n} -\epsilon_{b}(\vec{k}) }.
\ee
The cutoff function is designed to suppress the modes
with absolute value of band energy $|\epsilon_{b}(\vec{k})|$ below the scale $\Lambda$,
\be
C^\Lambda [\epsilon_{b}(\vec{k})] = \Theta_{\varepsilon} \bigl(
|\epsilon_{b}(\vec{k})| - \Lambda \bigr)\,.
\ee
We chose a smoothed step function $\Theta_{\varepsilon}$, where the width of the step is
characterized by a softening parameter $\varepsilon$. The modified
scale-dependent propagator $G_0^\Lambda$ gives rise to a modified bare action. Performing
the functional integral yields the effective action $\Gamma^\Lambda$, cf. Refs.~\onlinecite{wetterich1992,metzner2011,negorl}, which
serves as a generating functional of scale-dependent 1PI vertex functions.

By acting with $d/d\Lambda$ on $\Gamma^\Lambda$, one obtains an infinite hierarchy of coupled flow equations for the 1PI
vertex functions, in close analogy to the infinite tower of coupled Dyson-Schwinger equations. Integrating the flow down from some initial scale $\Lambda_0$ we smoothly interpolate between the bare action and the effective action at energy $\Lambda$. Correspondingly, we obtain 
the 1PI vertices of the effective action at scale $\Lambda$. The resulting flow
equations are valid for systems with $U(1)$ charge symmetry. Their recent use enabled
insights into correlated electron systems without spin-rotational symmetry, such as
correlated electron systems in the spin-orbit regime~\cite{scherer2014,schober2014} or in the presence of magnetic order\cite{maier2014}.

Here, we limit ourselves to the rather successful standard truncation of the infinite tower of differential
equations, suited for analyzing instabilities in correlated fermion systems. We keep only the flow of the static 4-point vertex,
i.e., the effective instantaneous interaction $V^{\Lambda}$. We neglect both self-energy feedback and the flow of
higher-order vertices which are generated upon running the mode elimination procedure
by integrating the flow equations. We will further justify this truncation in Sect.~\ref{sec:conclusions}.

The flow of the 4-point vertex is given by
\be
\frac{d}{d\Lambda} V^{\Lambda} = \phi_{\mathrm{pp}} + \phi_{\mathrm{ph,d}} - \phi_{\mathrm{ph,cr}},
\label{eq:flow}
\ee
where the particle-particle bubble $\phi_{\mathrm{pp}}$ and the crossed and direct particle-hole
bubbles $\phi_{\mathrm{ph,cr}}$ and $\phi_{\mathrm{ph,d}}$, respectively, are understood as bilinear functionals of the scale-dependent vertex function $V^{\Lambda}$. Fig.~\ref{fig:diagrams} shows a diagrammatic representation of the flow equation~\Eqref{eq:flow}. The initial condition $V^{\Lambda_{0}}$ for the vertex function is given by the coupling function $V_{b_1,b_2,b_3,b_4}(k_1,k_2,k_3,k_4)$ entering the Matsubara action. For explicit expressions of the flow equation~\Eqref{eq:flow}, see Appendix~\ref{app:flow}.

The renormalization of the vertex function obtained by solving \Eqref{eq:flow} numerically within a so-called patching scheme yields valuable information about the low-energy properties
of the model in question and indicates instabilities of metallic or semi-metallic phases towards symmetry-broken ground-states by a flow to strong coupling. The details of the patching discretization can be found in Fig.~\ref{fig:patching} and Appendix~\ref{app:patching}.

From the evolving pronounced momentum structure one can then infer the leading ordering tendencies. Since we find that the occurring instabilities are most easily interpreted in terms of the sublattice basis, after running the fRG flow, we finally transform the vertex function from the band to the sublattice basis, with the sublattice index $o=A, B$.

\section{Instability analysis}
\label{sec:instabilities}

We solve the fRG flow~\Eqref{eq:flow} for the effective interaction $V^{\Lambda}$ numerically, within the approximations described in Sect.~\ref{sec:frg}. We sweep through values of the couplings $V_{1}$ and $V_{2}$ in the parameter range $V_{1}/t$, $V_{2}/t \in [0,4]$. We note, however, that in the region in parameter space where either $V_{1} \gtrsim 2 t$ or $V_{2} \gtrsim 2 t$, the largest component of the vertex function fulfills $\max V^{\Lambda_{0}} \gtrsim  2 D$ with the bandwidth $D = 6 t$. In this region,
our weak coupling truncation of the fRG hierarchy is rendered unreliable.

\begin{figure}[t!]
\centering
\includegraphics[height=0.7\columnwidth]{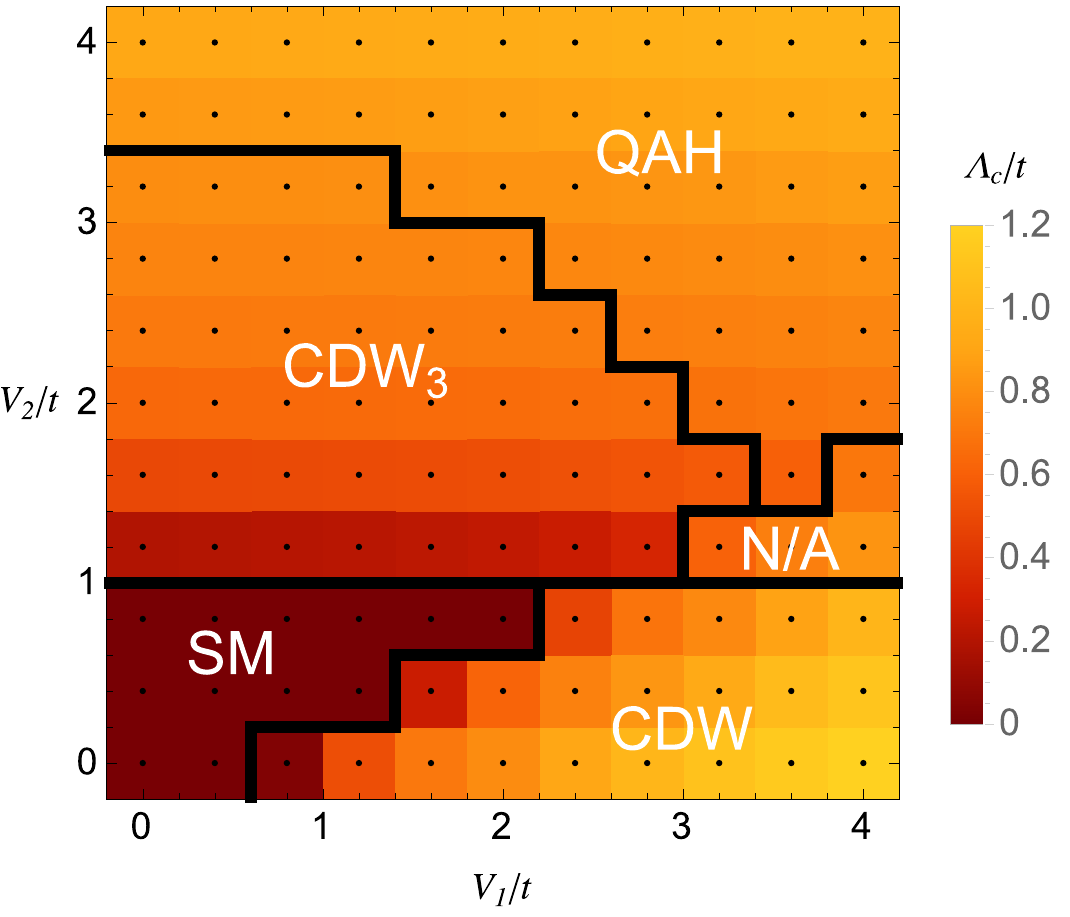}
\caption{Phase diagram for spinless fermions on the honeycomb lattice at half-filling in the plane of repulsive
nearest-neighbor and 2nd nearest-neighbor density-density interactions $V_{1}$ and $V_{2}$, respectively. The results were obtained with a $N=144$ patching-scheme, see Fig.~\ref{fig:patching}. The leading instability for $V_{1} = 0$, $V_{2} > V_{2,c}$ is the three-sublattice charge density wave (CDW$_{3}$). For $V_{1} > V_{1,c}$, $V_{2} = 0$, the leading instability is the conventional charge density wave (CDW) with charge imbalance between the $A$ and $B$ sublattices. The semi-metallic (SM) state shows an extended region of stability. The color-scale corresponds to the critical scales $\Lambda_{c}$ in units of the nearest-neighbor hopping $t$. For very large interactions, the quantum anomalous Hall (QAH) instability emerges in the phase diagram. This might be an artefact due to the breakdown of the weak coupling approximation. In the region marked with N/A, where the three different charge ordering instabilities meet, we observe strong $K$-$K$ scattering. A clear identification of the leading instability is this region was, however, not possible.}
\label{fig:phasediagram}
\end{figure}

In the case of a flow to strong coupling, only certain components of the momentum dependent vertex will diverge. The divergent momentum pattern, along with the corresponding sublattice configurations of the vertex function, yield an effective interaction functional. This in turn can be translated into an effective low-energy Hamiltonian, due to the static approximation for the vertex function. From the mean-field ground-state of the effective Hamiltonian, we obtain a tentative phase diagram for the spinless fermion model on the honeycomb in the $(V_{1},V_{2})$ parameter space. Our results are collected in the phase diagram depicted in Fig.~\ref{fig:phasediagram}, along with estimates for the typical energy scales, where long-range order starts building up. These are inferred from the renormalization group scale where the vertex starts to diverge. We denote the corresponding \textit{critical scale} as $\Lambda_{c}$. The values for $\Lambda_{c}$ can be regarded as estimates for critical temperatures.

\textit{Semi-metallic region.} We find an extended region of stability for the semi-metallic state. In this regime, the flow remains regular and no signs of a flow to strong coupling appear. Beyond certain values of $V_{1}$, $V_{2}$, the semi-metallic state shows instabilities in the particle-hole channel toward various charge-ordered states.

\begin{figure}[t!]
\begin{minipage}{1.0\columnwidth}
\centering
\includegraphics[width=1.0\columnwidth]{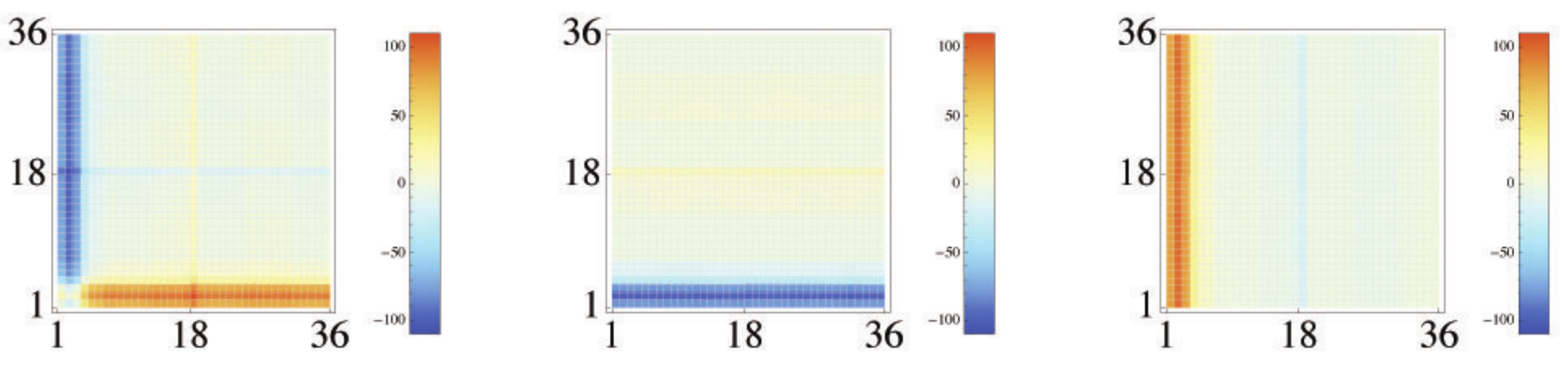}
\vspace{1.0em}
\end{minipage}
\begin{minipage}{1.0\columnwidth}
\centering
\includegraphics[width=1.0\columnwidth]{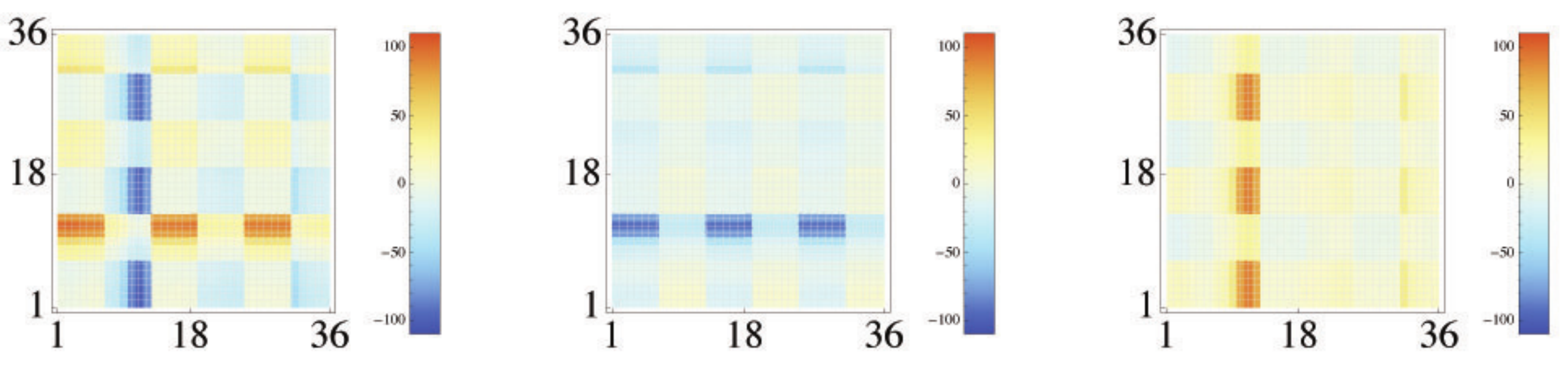}
\end{minipage}
\caption{Upper panel: Vertex structure in sublattice representation for divergent CDW correlations for $V_{1} = 1.2 t$, $V_{2}= 0$ at the critical scale $\Lambda_{c} \approx 0.26 t$ computed in $N=144$ patching scheme as depicted in Fig.~\ref{fig:patching}. From left to right: $V_{AAAA}^{\Lambda_{\mathrm{c}}}$, $V_{ABAB}^{\Lambda_{c}}$, $V_{ABBA}^{\Lambda_{c}}$. The patch number of the patch momentum $\pi(\vec{k}_{1})$ is on the vertical, the patch number of $\pi(\vec{k}_{2})$ on the horizontal axis, while $\pi(\vec{k}_{3})$ is fixed to the second patch (see. Fig.~\ref{fig:patching}). For clarity, we only display the vertex for the $36$ patches closest to the $K$, $K^{\prime}$ points. The divergent momentum structure can be translated to an effective Hamiltonian \Eqref{eq:HCDW} favoring a CDW ground state.
Lower panel: Vertex structure in sublattice representation for divergent CDW$_{3}$ correlations for $V_{1} = 0 $, $V_{2}= 1.2 t$ at the critical scale $\Lambda_{c} = 0.70 t$ computed in $N=144$ patching scheme as depicted in Fig.~\ref{fig:patching}, with the same conventions as above. The divergent momentum structure corresponds to enhanced scattering between fermions with a momentum transfer $\vec{Q} \approx \vec{K} - \vec{K}^{\prime}$. This translates to the effective Hamiltonian \Eqref{eq:HCDW3} with a CDW$_{3}$ ground state. The faint checkerboard-like pattern in the background resembles the initial condition of the vertex function. During a portion of the flow, this pattern is also enhanced as a whole.}
\label{fig:vertexCDWs}
\end{figure}

\begin{figure}[t!]
\centering
\includegraphics[height=0.4\columnwidth]{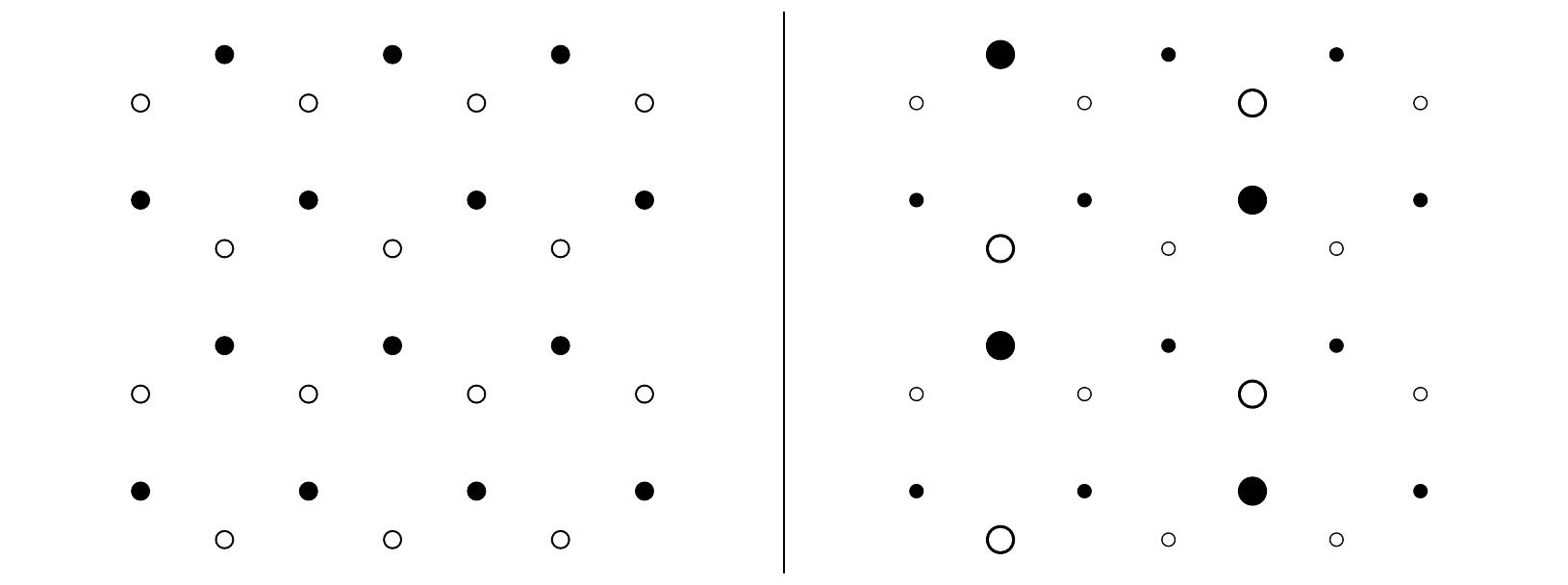}
\caption{Density modulations of CDW and CDW$_{3}$ patterns. White circles denote lattice
sites with decreased density, while black circles correspond to sites with increased density. The radius is a measure
for the local deviation from the average charge density. Left: The conventional CDW pattern preferring the occupation of e.g. the $B$ sublattice. Right: An example of a CDW$_{3}$ pattern with $\alpha=\pi/3$. Around one plaquette of the honeycomb, the charge modulation on each site differs, corresponding to 
the breaking of each sublattice into three new sublattices. Looking at the plaquette in the lower right corner, in counter clockwise direction, we obtain a charge modulation of $(-\delta, +\delta, -\delta, +\delta, -\Delta, + \Delta)$, where $\delta$ and $\Delta$ refer to the local charge imbalance and $\Delta > \delta$.}
\label{fig:honeyCMs}
\end{figure}

\textit{CDW instability.} Let us consider the case $V_{2} = 0$ first. Beyond a critical coupling $V_{1,c}/t \approx 0.6$, the SM state is destabilized, and we encounter a CDW instability. The diverging momentum structure for the patch-points closest to $K, K^{\prime}$ is depicted in the upper row of Fig.~\ref{fig:vertexCDWs}. The CDW instability is driven by scattering of states with zero momentum transfer. There is no modulation  of the scattering amplitude, as the momentum values of the incoming and outgoing states sweep through the representative patch momenta. The effective interaction Hamiltonian close to the CDW instability extracted from the numerical data for the renormalized vertex reads
\be
H_{\mathrm{eff}}^{\Lambda_{\mathrm{c}}} = - \frac{1}{\mathcal{N}} \sum_{o,o^{\prime}} V_{o,o^{\prime}}\epsilon_{o}\epsilon_{o^{\prime}} N_{\vec{q}=0}^{o} \, N_{\vec{q}=0}^{o^{\prime}},
\label{eq:HCDW}
\ee
where $V_{o,o^{\prime}} > 0$, and $\mathcal{N}$ is the number of unit cells. The prefactors $\epsilon_{A} = +1$, $\epsilon_{B} = -1$ capture the sublattice modulation. The Hamiltonian obviously factorizes into a product of density operators $N_{\vec{q}}^{o} = \sum_{\vec{k}} c^{\dagger}_{o, \vec{k}+\vec{q}}c_{o, \vec{k}}$ for $\vec{q}=0$. This translates into an infinitely ranged density-density interaction on the lattice. The sublattice modulation prefers the occupancy of either the $A$ or the $B$ sublattice through both attractive intra-sublattice components $\sim - N_{\vec{q}=0}^{A} \, N_{\vec{q}=0}^{A}$, $\sim - N_{\vec{q}=0}^{B} \, N_{\vec{q}=0}^{B}$, as well as a repulsive inter-sublattice component $\sim \left(N_{\vec{q}=0}^{A} \, N_{\vec{q}=0}^{B} + N_{\vec{q}=0}^{B} \, N_{\vec{q}=0}^{A}\right)$. Since the effective low-energy Hamiltonian (unintegrated part of valence and conduction bands + effective interaction) conserves particle number, the resulting ground-state will
have either the $A$ or $B$ sublattice occupied/empty. This corresponds to the spontaneous breaking of the discrete sublattice symmetry, see Fig.~\ref{fig:honeyCMs} for 
the resulting sublattice occupation. We note that the diverging momentum pattern with momenta on the 1st patch rings closest to the $K$, $K^{\prime}$ points appears slightly blurred, because in our discretization the $K$-$K$ scattering is represented by all momenta on the first patch rings. The momentum pattern on higher patch rings, however, becomes very sharp.

\textit{CDW$_{3}$ instability.}
We now consider the case $V_{1} = 0$. For strong 2nd neighbor coupling, the semi-metallic state turns unstable at a critical coupling of $V_{2,c} \approx 1.0 t$. It does not, however, correspond to a QAH instability, as one would expect based on previous mean-field results~\cite{raghu2008,weeks2010,dauphin2012,grushin2013}. The emerging instability has a rather peculiar momentum structure. We show the diverging momentum structure in the lower row of Fig.~\ref{fig:vertexCDWs}. Also here, the momentum structure for momenta on patch-rings further away from the $K$, $K^{\prime}$ points is very sharp. We find a striking $\vec{Q} \neq 0$ signature, indicating the tendency toward the formation of unit-cell enlarging order. The intra-sublattice component $V_{AAAA}^{\Lambda_{c}}$ of the interaction vertex, for example, is dominated by outgoing momenta $\vec{k}_{1}$, $\vec{k}_{2} = \vec{k}_{3} + \vec{Q}$ (vertical feature) and $\vec{k}_{1} = \vec{k}_{3} + \vec{Q}$, $\vec{k}_{2}$ (horizontal feature). This, and also the intermittent amplitude of the interaction vertex along these features, correspond to enhanced $K$--$K^{\prime}$ scattering with a momentum transfer of $\vec{Q} \approx \vec{K}-\vec{K}^{\prime}$. With the same definitions as above, we extract the effective interaction Hamiltonian as
\be
H_{\mathrm{eff}}^{\Lambda_{\mathrm{c}}} = - \frac{1}{\mathcal{N}} \sum_{o,o^{\prime}} V_{o,o^{\prime}}\epsilon_{o}\epsilon_{o^{\prime}} \left(
N_{\vec{Q}}^{o} \, N_{-\vec{Q}}^{o^{\prime}} + 
N_{-\vec{Q}}^{o} \, N_{\vec{Q}}^{o^{\prime}}
\right).
\label{eq:HCDW3}
\ee
Upon transforming to real space, one arrives again at an infinitely ranged interaction. But in the present case, the amplitude shows a modulation with the wavevector $\vec{Q}$. An analogous effective interaction was already obtained for an extended Hubbard model on honeycomb bi- and trilayers~\cite{mmscherer2012,mmscherer2012b,delapena2014}. Since the interaction in \Eqref{eq:HCDW3} factorizes into a product of fermion bilinears $N_{\vec{Q}}$ and is infinitely ranged, mean-field theory is expected to yield reasonably accurate information about the ground-state. It turns out that in self-consistent mean-field approach, the energy is minimized by a finite complex order parameter $\langle N_{\vec{Q}} \rangle = \epsilon_{o}\Delta_{o}\mathrm{e}^{\mathrm{i}\alpha}$, described by its amplitude $\Delta_{o}$ and phase $\alpha$. 
As observed already in Ref.~\onlinecite{mmscherer2012}, this gives rise to a density modulation $\sim \cos\left(\vec{Q}\cdot\vec{R}+\alpha\right)$ on the honeycomb lattice, and a concomitant 6-atom unit cell, cf. Fig.~\ref{fig:honeyCMs}. This implies that each sublattice $A$, $B$ is broken up into three new sublattices $A_{1}$, $A_{2}$, $A_{3}$ and $B_{1}$, $B_{2}$, $B_{3}$. For this reason, the notion of a CDW$_{3}$ was coined in Ref.~\onlinecite{mmscherer2012} for this three-sublattice charge-density wave. The phase parameter $\alpha$ describes the redistribution of charge in each of the emergent sublattices, while the average remains constant upon changing $\alpha$. The ground-state energy is minimized for $\alpha = n \pi /3$ with $n$ integer. A state with $\langle N_{\vec{Q}} \rangle \neq 0$ and $\alpha = n \pi /3$ opens a charge gap, in agreement with previous results~\cite{grushin2013,garcia2013,daghofer2014}.

We note, however, that not only the components of the vertex function with momenta close to the $K, K^{\prime}$ points flow to large values. Also the momenta further away from the BZ corners grow during the flow, and show strong $\vec{Q} \neq 0$ signatures. In fact, we find that the inclusion of these momenta is crucial in obtaining the CDW$_{3}$ as the leading instability close to the semi-metallic regime. 

The QAH instability and its Hamiltonian will be discussed in Sect.~\ref{sec:suppression}. We note that for very large $V_{2}$, we observe flows with both strong CDW$_{3}$ and QAH signatures. For increasing $V_{2}$, the QAH features will eventually dominate, at least on the first and second patch ring. Since the fRG in the present truncation is a weak-coupling method, the appearance of the QAH instability in this parameter regime is possibly an artefact of the breakdown of the weak-coupling approximation.

\section{Suppression of interaction induced topological insulator phase}
\label{sec:suppression}

Within previous standard patching schemes for the analysis of Fermi surface instabilities~\cite{metzner2011} in correlated fermion systems, the momentum dependence of the vertex function was typically projected to patch momenta located on the Fermi surface. For nodal fermionic systems, where the non-interacting degrees of freedom are characterized by a vanishing density of states at the Fermi level, this approximation might not capture all relevant contributions to the renormalization of the interaction vertex. In the following, we restrict our attention to $V_{1} =0$, $V_{2} > 0$.
\begin{figure}[t!]
\centering
\includegraphics[height=0.225\columnwidth]{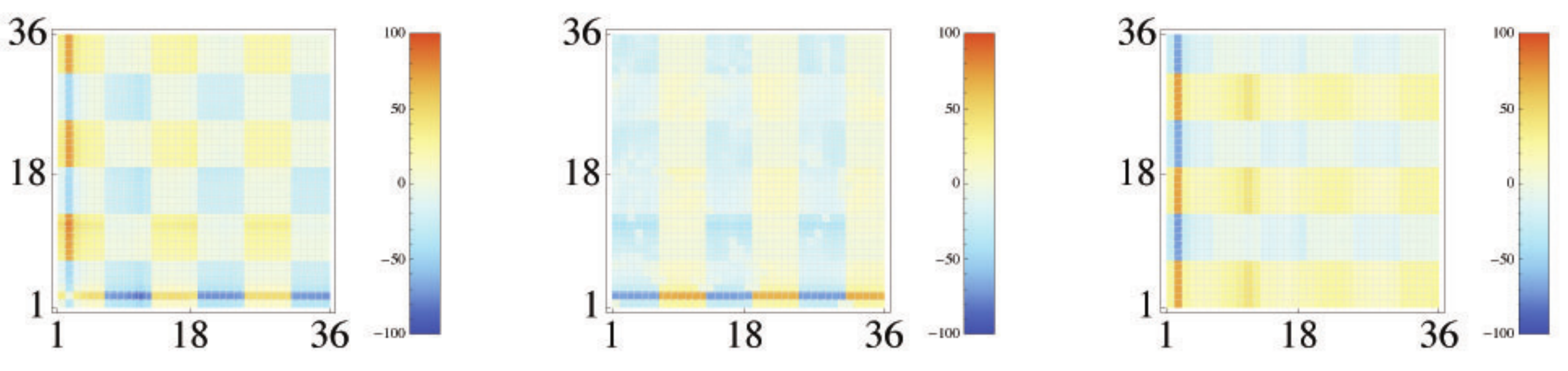}
\caption{Vertex structure in sublattice representation for divergent QAH correlations for $V_{1} = 0 $, $V_{2}= 0.8 t$ close to the critical scale $\Lambda_{c} \approx 0.31 t$ computed in a $N = 36$ patching scheme. From left to right: $V_{AAAA}^{\Lambda_{c}}$, $V_{ABAB}^{\Lambda_{c}}$, $V_{ABBA}^{\Lambda_{c}}$. The sharp horizontal and vertical features display the $f$-wave amplitude modulation, see \Eqref{eq:HQAH}.
The number of the patch momentum $\pi(\vec{k}_{1})$ is enumerated on the vertical, the number of $\pi(\vec{k}_{2})$ on the horizontal axis, while $\pi(\vec{k}_{3})$ is fixed to the second patch. The checkerboard pattern in the background, which resembles the form of the initial condition of the vertex function, also grows to sizable values. The characteristic CDW$_{3}$ signature with a momentum transfer $\vec{Q}\approx \vec{K} - \vec{K}^{\prime}$ is barely visible as a subleading momentum pattern.}
\label{fig:vertexQAH}
\end{figure}

For the spinless fermion model on the honeycomb lattice, the simplest patching scheme resolves only the momenta closest to the $K$, $K^{\prime}$ points,  i.e., only the first patch ring is taken into account, cf. Fig.~\ref{fig:patching}. This scheme resolves the non-trivial angular dependence of orbital make-up, but neglects the radial momentum dependence of the vertex function away from the $K$, $K^{\prime}$ points. Running the fRG flow (with e.g. $N=36$ patches), we obtain a momentum signature as depicted in Fig.~\ref{fig:vertexQAH}. The effective interaction Hamiltonian turns out as
\be
H_{\mathrm{eff}}^{\Lambda_{\mathrm{c}}} = - \frac{1}{\mathcal{N}} \sum_{o,o^{\prime}} V_{o,o^{\prime}}\epsilon_{o}\epsilon_{o^{\prime}} S_{f,\vec{q}=0}^{o} \, S_{f,\vec{q}=0}^{o^{\prime}},
\label{eq:HQAH}
\ee
with the operator $S_{f,\vec{q}}^{o} = \sum_{\vec{k}} f_{\vec{k}}c^{\dagger}_{o, \vec{k}+\vec{q}}c_{o, \vec{k}}$ and a form factor $f_{\vec{k}} = \sin(\sqrt{3}k_{x}) - 2 \sin(\sqrt{3}k_{x}/2)\cos(3k_{y}/2)$.
The variational ground-state of the Hamiltonian~\Eqref{eq:HQAH} is given by a purely imaginary $\langle S_{f,\vec{q}=0}^{o} \rangle$. Moving to real space, this corresponds to a finite, purely imaginary dimerization amplitude $\chi_{ij}$ on 2nd neighbor bonds, i.e., the QAH state.

\begin{figure}[t!]
\begin{minipage}{0.45\columnwidth}
\centering
\includegraphics[width=1.0\columnwidth]{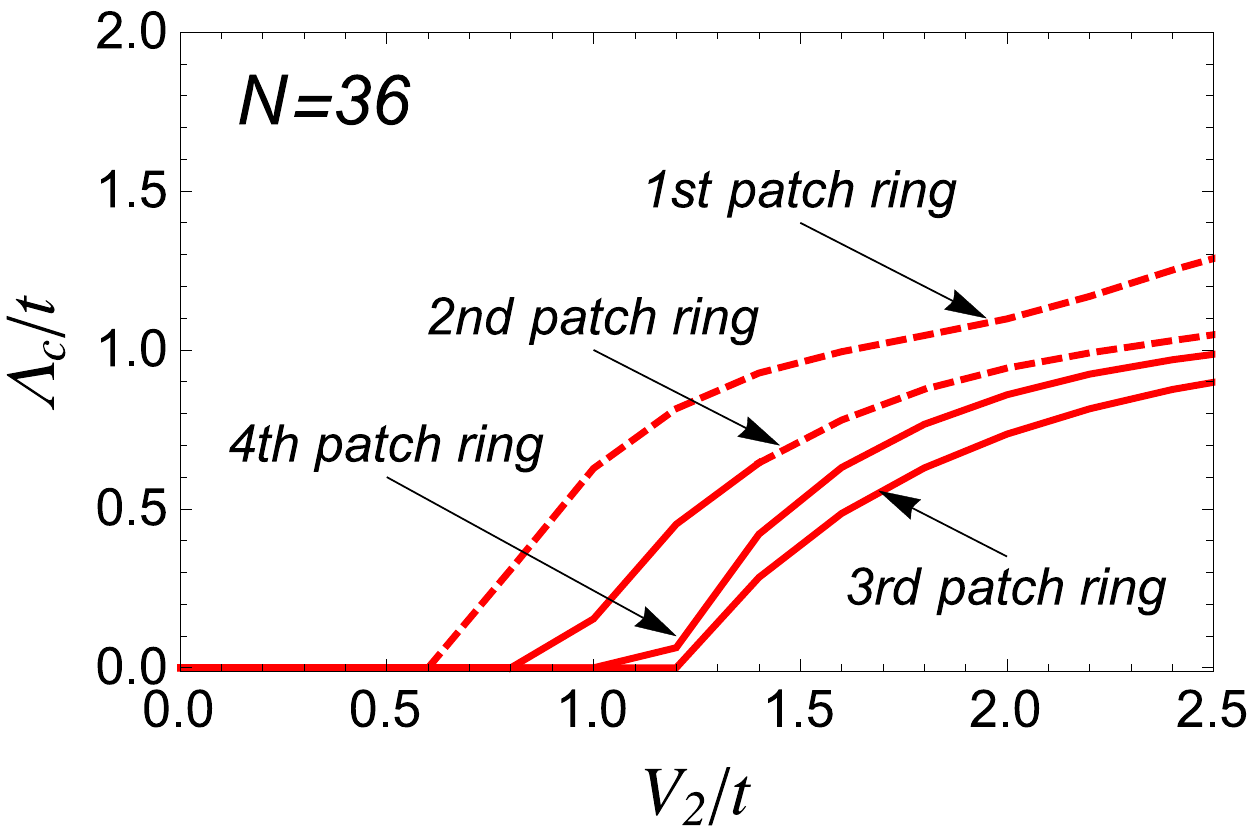}
\end{minipage}
\hspace{0.5em}
\begin{minipage}{0.45\columnwidth}
\centering
\includegraphics[width=1.0\columnwidth]{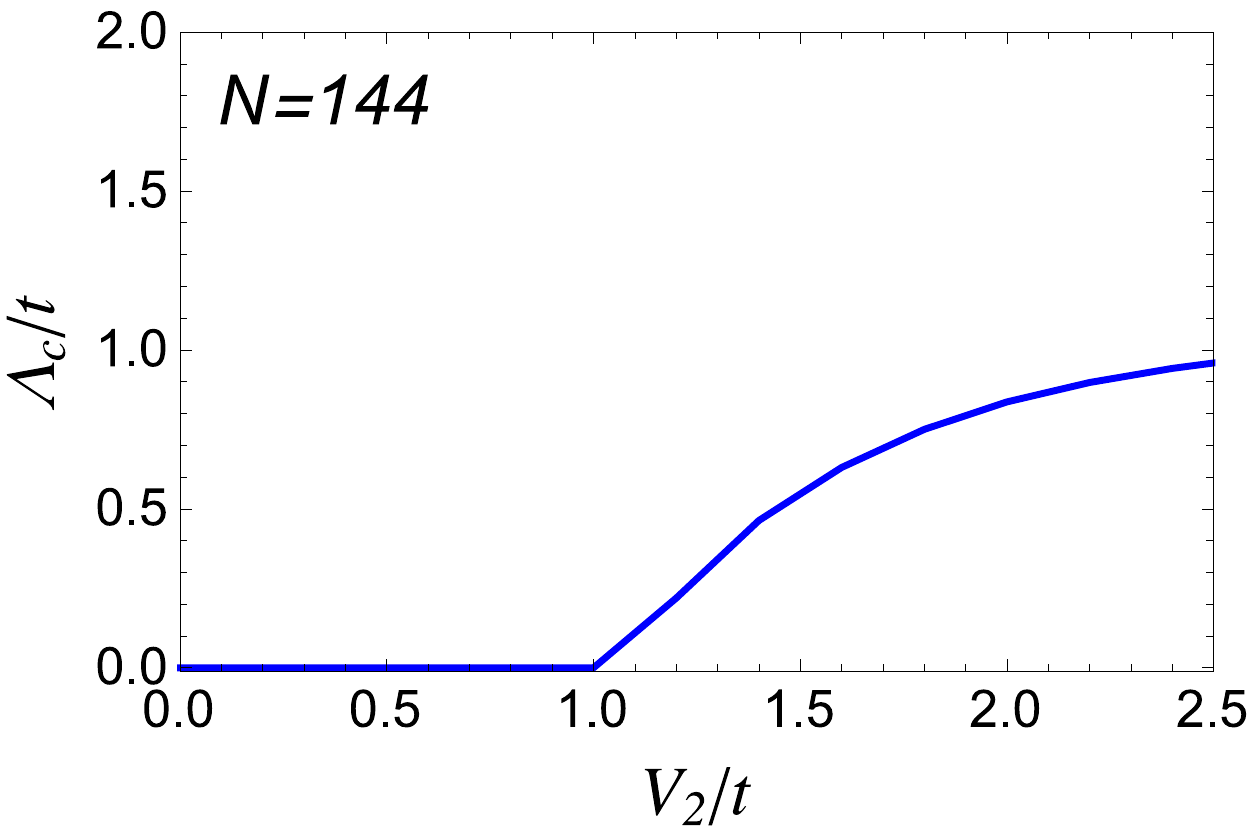}
\end{minipage}
\caption{Red curves: Evolution of critical scales $\Lambda_{c}/t$ from a $N=36$ patching as a function of $V_{2}/t$ for $V_{1} = 0$, as the patch-momenta are shifted closer to the $\Gamma$ point. The curves are labeled according to the patch ring (cf. Fig.~\ref{fig:patching}) entering the approximation. The dashing indicates, when the QAH takes over the CDW$_{3}$ as the leading instability. While the patching with only the second patch ring is susceptible to detecting the CDW$_{3}$ in a small parameter range, the patching with the third patch ring only yields the CDW$_{3}$ over the full parameter range. The fourth patch ring yields higher critical scales than the third again, indicating the importance of momenta further away from the BZ corners. Blue curve: Critical scales $\Lambda_{c}/t$ as a function of $V_{2}/t$ for $V_{1} = 0$ within the $N=144$ patching scheme with all momentum configurations of the vertex coupled to each other. Apparently, the blue curve is a compromise of the critical scales obtained with the one-patch-ring approximation.}
\label{fig:critscales}
\end{figure}

Within this approximation to the vertex, the CDW$_{3}$ can be identified 
as a subleading momentum pattern for certain values of $V_{2}$ close to $V_{2,c}$. To analyze the effect of shifting the patch momenta to higher patch rings further away from the $K$, $K^{\prime}$ points (cf. Fig.~\ref{fig:patching}) on the approximation, we keep $N=36$ patches, but move the position of the representative patch momenta closer to the $\Gamma $ point and run the fRG flow for the vertex function projected onto the new set of patch momenta. The critical scales as a function of $V_{2}$ corresponding to these (arguably unphysical approximations) are collected in Fig.~\ref{fig:critscales}. We observe that as we move the representative momenta to higher patch rings, the critical scales tend to move down, and correspondingly, the value of the critical coupling shifts to larger values. Eventually, the CDW$_{3}$ takes over the QAH instability in the full range of $V_{2}$ values, as we project the momentum dependence of the vertex on the third patch ring (counting from the BZ corners).

Including all $N=144$ patch momenta correpsonding to our best momentum resolution as depicted in Fig.~\ref{fig:patching}, we obtain the blue curve in Fig.~\ref{fig:critscales} for the critical scales. The CDW$_{3}$ is found as the leading instability for $V_{2} < 3.4 t$. It is justified to say that by including momentum configurations with external legs on any of the patch rings, and allowing all configurations to talk to each other, the QAH instability is in fact suppressed over a large portion of parameter space. This is further evidenced by the fact that upon inclusion of all patch rings, the critical scales actually move down, compared to the approximations involving only a single patch ring. If no suppression of the QAH signature were at work, it would simply diverge, even before the CDW$_{3}$ signature has cascaded down from the highest to the lowest patch ring.

We further checked our results within another discretization scheme, which resolves also momenta close to the $\Gamma$ point. We find the same qualitative behaviour, and only small quantitative corrections in the value of the critical coupling to larger and the critical scales to smaller values.

We note that the CDW instability remains largely unaffected upon changing the resolution of our discretization.

\section{Conclusions \& Discussion}
\label{sec:conclusions}

We have to investigated the phase diagram of spinless fermions on the honeycomb lattice at half-filling with repulsive nearest and 2nd nearest-neighbor interactions by the functional renormalization group. The fRG represents a modern implementation of the Wilsonian renormalization group idea of successively eliminating high-energy degrees of freedom, while tracking the evolution of effective couplings in the effective low-energy theory. We compute the flow of the momentum-dependent effective interaction (4-point vertex) without self-energy feedback. This corresponds to an unbiased resummation of 1-loop diagrams in both particle-particle and particle-hole channels contributing to the effective interaction. Performing an instability analysis, we find CDW and CDW$_{3}$ instabilities and a suppression of the QAH topological Mott insulator.

Our fRG approach is a suitable tool for the purpose of checking the ground-state manifold of the model for the presence/absence of the QAH phase in the parameter regime where $\max V^{\Lambda_{0}} \lesssim 2 D$ holds for the following reasons.
(i) While mean-field theories with extended unit cells~\cite{jia2013,grushin2013} detect both QAH and modulated CDW phases, the absence of QAH in other mean-field calculations and in particular exact diagonalization~\cite{garcia2013,daghofer2014,capponi2015} and iDMRG~\cite{motruk2015} points to a competing instability scenario. The fRG, and in particular the `working-horse' truncation described in Sect.~\ref{sec:frg}, was shown to capture the competition of different ordering tendencies for low-dimensional fermion models~\cite{metzner2011} quite reliably. 
(ii) Dropping self-energy feedback and the flow of the 6-point vertex and higher order vertex functions seems justified for the purpose we have in mind, as long as the coupling $V_{2}$ remains sufficiently small. While a partial inclusion of higher-order flow-diagrams in the flow of the effective interaction captures the effect of collective fluctuations, this kind of truncation is only necessary, when one is after realistic gap sizes and critical temperatures or an accurate description of critical behavior in proximity to a QCP with gapless collective excitations. As mentioned already above, in the present case, the instability in question (QAH) corresponds to breaking of a discrete symmetry, i.e., collective fluctuation effects are expected to play a minor role for the competition of QAH and modulated CDW instabilities. Thus, the truncation described in Sect.~\ref{sec:frg} should suffice to correctly identify the fermionic fluctuations driving the system critical.

In summary, we presented strong evidence for the presence of a direct transition from the SM to the CDW$_{3}$, without an intervening topological Mott insulator state. For very large $V_2$, however, from our fRG results we cannot rule out that the QAH phase may be stable. Since several numerical approaches~\cite{jia2013,daghofer2014, capponi2015,motruk2015} yield phase diagrams without any sign of a QAH phase, the dominance of the QAH instability over the CDW$_3$ in our fRG flows for very large couplings is most probably caused by the failure of our truncation to capture strong-coupling effects.
As the bare interaction strengths grow larger, the relevant physics can be captured by an Ising-type model~\cite{capponi2015,motruk2015}, where the hopping acts as a perturbation. At stronger coupling, a real-space formulation of the fRG, similar in spirit to spin-fRG~\cite{spinfRG}, could provide a more reliable starting point in obtaining an accurate phase diagram in the strong coupling regime, than our present formulation based on itinerant degrees of freedom. Further, our weak coupling study did not detect any signs of a Kekul$\acute{\textrm{e}}$ bond-order instability or charge-order instabilities with quadrupled unit cells as reported in Refs.~\onlinecite{capponi2015,motruk2015} in the parameter range considered. Within the weak-coupling regime of the model, we find qualitative agreement with recent numerical studies~\cite{jia2013,daghofer2014, capponi2015,motruk2015}.

Some other interesting questions also cannot be easily accessed within our truncation, such as the order of the transition line
between CDW and CDW$_{3}$, or the properties of the corresponding multicritical point. While the QCP of the CDW order
is well investigated (see Sect.~\ref{sec:intro}), much less appears to be known about the QCP corresponding to CDW$_{3}$ order and the
corresponding effective field theory.  

The case away from half-filling was studied e.g. in Reference~\onlinecite{grushin2013} by self-consistent mean-field theory.
Our fRG results for the spinless fermion model on the honeycomb at chemical potential $\mu \neq 0 $ will be presented in a forthcoming paper.

The authors are grateful to S.~E.~Seidenbecher and I.~Boettcher for critical comments on the manuscript.

\appendix

\section{Flow Equations}
\label{app:flow}

In shortened notation, the bubble expressions are given by
\be
\phi_{\mathrm{pp}} & = & \frac{1}{2}V^{\Lambda}\circ L\circ V^{\Lambda}|_{\mathrm{pp}}, \\
\phi_{\mathrm{ph,d}} & = & -\frac{1}{4}V^{\Lambda}\circ L\circ V^{\Lambda}|_{\mathrm{ph,d}}, \\
\phi_{\mathrm{ph,cr}} & = & -\frac{1}{4}V^{\Lambda}\circ L\circ V^{\Lambda}|_{\mathrm{ph,cr}},
\ee
where the $\circ$ symbol denotes the channel specific contractions between loop Kernel $L$
and vertex functions $V^{\Lambda}$. In our approximation, the loop Kernel $L = S^{\Lambda} G_{0}^{\Lambda} + G_{0}^{\Lambda} S^{\Lambda}$ is built from the bare scale-dependent propagator $G_{0}^{\Lambda}$ and the so-called single-scale propagator $S^{\Lambda} = - d/d\Lambda\, G_{0}^{\Lambda}$. The explicit expressions for particle-particle and direct particle-hole bubbles read
\be
\phi_{\mathrm{pp}}(X_1,\xi_2,\xi_3,\xi_4) = \frac{1}{2}\prod_{\nu=1}^{4}\!\int\!\!d \eta_{\nu}\, L(\eta_2,\eta_1,\eta_3,\eta_4)\times \nn \\
V^{\Lambda}(\xi_2,\xi_1,\eta_2,\eta_3)V^{\Lambda}(\eta_4,\eta_1,\xi_3,\xi_4), \nn
\ee
\be
\phi_{\mathrm{ph,d}}(\xi_1,\xi_2,\xi_3,\xi_4)  =   - \frac{1}{4}\prod_{\nu=1}^{4}\!\int\!\!d \eta_{\nu}\, L(\eta_1,\eta_2,\eta_3,\eta_4)\times \nn \\
V^{\Lambda}(\eta_4,\xi_2,\xi_3,\eta_1)V^{\Lambda}(\xi_1,\eta_2,\eta_3,\xi_4), \nn
\ee
and the crossed particle-hole contribution is given through
\be
\phi_{\mathrm{ph,cr}}(\xi_1,\xi_2,\xi_3,\xi_4)  =  \phi_{\mathrm{ph,d}}(\xi_1,\xi_2,\xi_4,\xi_3).
\ee
Above, we introduced the shorthand $\int\!\!d \eta$ to represent integration/summation over
loop variables. In the band representation, $\xi = (b,\mathrm{i}\omega_{n},\vec{k})$, $\eta=(b^{\prime},\mathrm{i}\omega_{n}^{\prime},\vec{k}^{\prime})$. Since we are interested in ground-state properties, we make a static approximation for the vertex function and neglect frequency
dependence. 

\section{Projection onto Patch Momenta}
\label{app:patching}

The wavevector dependence of the interaction vertex is approximated in a so-called $N$-patch scheme. The Brillouin zone (BZ) is divided into $N$ patches. A given wavevector $\vec{k} \in$ BZ is projected onto the closest representative patch momentum, $\pi(\vec{k})$. See Fig.~\ref{fig:patching} for our patching discretization. A single patch is thus composed of all the momenta, which have the smallest Euclidean distance to the corresponding representative patch momentum. We then solve \Eqref{eq:flow} for the projected vertex function $V^{\Lambda}(\pi(\vec{k}_{1}),\pi(\vec{k}_{2}),\pi(\vec{k}_{3}),\pi(\vec{k}_{4}))$. We note that $V^{\Lambda}$ also depends on band indices $b=v, c$ of outgoing (first two arguments) and incoming (last two arguments) legs. 

\newpage




\begin{thebibliography}{n}

\bibitem{raghu2008}
S. Raghu, Xiao-Liang Qi, C. Honerkamp, and Shou-Cheng Zhang, Phys. Rev. Lett. {\bf 100}, 156401 (2008).

\bibitem{herbut2006}
I. F. Herbut, Phys. Rev. Lett. {\bf 97}, 146401 (2006).

\bibitem{herbut2009}
I. F. Herbut, V. Juri\v ci\'c, and Bitan Roy, Phys. Rev. B {\bf 79}, 085116 (2009).

\bibitem{rosa2001}
L. Rosa, P. Vitale, C. Wetterich, Phys. Rev. Lett. {\bf 86}, 958 (2001).

\bibitem{hofling2002}
F. H{\"o}fling, C. Nowak, C. Wetterich, Phys. Rev. B {\bf 66}, 205111 (2002).

\bibitem{braun2011}
J. Braun, H. Gies, D. D. Scherer, Phys. Rev. D {\bf 83}, 085012 (2011).

\bibitem{gross1974}
D.J. Gross and A. Neveu, Phys. Rev. D {\bf 10}, 3235 (1974).

\bibitem{rosenstein1988}
B. Rosenstein, B. J. Warr and S. H. Park, Phys. Rev. Lett. {\bf 62}, 1433 (1989).

\bibitem{rosenstein1993}
B. Rosenstein, H.-L. Yu, A. Kovner, Phys. Lett. {\bf B314}, 381 (1993)

\bibitem{janssen2014}
L. Janssen and I. F. Herbut, Phys. Rev. B {\bf 89}, 205403 (2014).

\bibitem{classen2015}
L. Classen, I. F. Herbut, L. Janssen, M. M. Scherer, Phys. Rev. B {\bf 92}, 035429 (2015).

\bibitem{khveshchenko2001}
D. V. Khveshchenko, Phys. Rev. Lett. {\bf 87}, 246802 (2001).

\bibitem{wang2014}
L. Wang, P. Corboz, M. Troyer, New J. Phys. {\bf 16}, 103008 (2014).

\bibitem{li2015A}
Zi-Xiang Li, Yi-Fan Jiang, and Hong Yao, Phys. Rev. B {\bf 91}, 241117(R) (2015).

\bibitem{li2015B}
Zi-Xiang Li, Yi-Fan Jiang, Hong Yao, arXiv:1411.7383 [cond-mat.str-el].

\bibitem{weeks2010}
C. Weeks and M. Franz, Phys. Rev. B {\bf 81}, 085105 (2010).

\bibitem{dauphin2012}
A. Dauphin, M. M\"uller, M.A. Martin-Delgado, Phys. Rev. A {\bf 86}, 053618 (2012).

\bibitem{grushin2013}
A. G. Grushin et al., Phys. Rev. B {\bf 87}, 085136 (2013).

\bibitem{haldane1988}
F. D. M. Haldane, Phys. Rev. Lett. {\bf 61}, 2015 (1988).

\bibitem{ryu2010}
S. Ryu, A. Schnyder, A. Furusaki, A. Ludwig, New J. Phys. {\bf 12}, 065010 (2010).

\bibitem{jia2013}
Y. Jia, H. Guo, Z. Chen, S.-Q. Shen, and S. Feng, Phys. Rev. B {\bf 88}, 075101 (2013).

\bibitem{daghofer2014}
M. Daghofer and M. Hohenadler, Phys. Rev. B {\bf 89}, 035103 (2014).

\bibitem{capponi2015}
S. Capponi, A. M. L\"auchli, arXiv:1505.01674 [cond-mat.str-el].

\bibitem{motruk2015}
J. Motruk, A, G. Grushin, F. de Juan, F. Pollmann, arXiv:1505.01676 [cond-mat.str-el].

\bibitem{mmscherer2012}
M. M. Scherer, S. Uebelacker, C. Honerkamp, Phys. Rev. B {\bf 85}, 235408 (2012).

\bibitem{mmscherer2012b}
M. M. Scherer, S. Uebelacker, D. D. Scherer, and C. Honerkamp,
Phys. Rev. B {\bf 86}, 155415 (2012).

\bibitem{garcia2013}
N. A. Garc\'ia-Mart\'inez et al., Phys. Rev. B {\bf 88}, 245123 (2013).

\bibitem{duric2014}
T. \DJ uri\'c, N. Chancellor, I. F. Herbut, Phys. Rev. B {\bf 89}, 165123 (2014).

\bibitem{negorl} 
J.~W.~Negele, H.~Orland, {\em Quantum many-particle systems}, Addison-Wesley Publishing Company (1988).

\bibitem{kopietz2010}
P. Kopietz, L. Bartosch, and F.~Sch{\"u}tz, \emph{Introduction to the
functional renormalization group}, Springer Verlag, Berlin (2010).

\bibitem{metzner2011} 
W.~Metzner, M.~Salmhofer, C.~Honerkamp, V.~Meden, and Kurt Sch\"onhammer, Rev. Mod. Phys. {\bf 84}, 299 (2012).

\bibitem{platt2013}
C.~Platt, W.~Hanke, R.~Thomale, Advances in Physics, Volume {\bf 62}, Issue 4-6, 2013.

\bibitem{wetterich1992} 
  C.~Wetterich, Phys.\ Lett.\ B {\bf 301}, 90 (1993).

\bibitem{maier2013}
S. A. Maier, J. Ortloff, and C. Honerkamp, Phys. Rev. B 88, 235112 (2013).

\bibitem{scherer2014}
D. D. Scherer, M. M. Scherer, G. Khaliullin, C. Honerkamp, and B. Rosenow, Phys. Rev. B {\bf 90}, 045135 (2014).

\bibitem{schober2014}
G. A. H. Schober, K.-U. Giering, M. M. Scherer, C. Honerkamp, M. Salmhofer,  arXiv:1409.7087.

\bibitem{maier2014}
S. A. Maier, A. Eberlein, and C. Honerkamp, Phys. Rev. B {\bf 90} 035140 (2014).

\bibitem{delapena2014}
S{\'a}nchez de la Pe{\~n}a, D., Scherer, M.~M., 
\& Honerkamp, C.\ Ann. Phys., 526: 366-371, arXiv:1407.5582

\bibitem{spinfRG}
J.~Reuther, P.~W\"{o}lfle, Phys. Rev. B {\bf 81}, 144410 (2010); J.~Reuther, R.~Thomale, Phys. Rev. B {\bf 83}, 024402 (2011); J.~Reuther {\it et al.}, Phys. Rev. B {\bf 83}, 064416 (2011); Yogesh~Singh {\it et al.}, Phys. Rev. Lett. {\bf 108}, 127203 (2012);
J.~Reuther, R.~Thomale, and S.~Rachel, Phys. Rev. B {\bf 86}, 155127 (2012); R.~Suttner, C.~Platt, J.~Reuther, R.~Thomale, arXiv:1303.0579; J.~Reuther, R.~Thomale, arXiv:1309.3262;
\end{thebibliography}
\end{document}